\documentclass[conference]{IEEEtran}
\usepackage[hyphens]{url}
\usepackage{hyperref}
\hypersetup{colorlinks,allcolors=blue}

\usepackage[backend=bibtex,style=ieee,citestyle=numeric-comp]{biblatex}
\usepackage{amsmath,amssymb,amsfonts}
\usepackage{algorithmic}
\usepackage{graphicx}
\usepackage{textcomp}
\usepackage{xcolor}
\usepackage{comment}
\usepackage{enumitem} 
\usepackage{orcidlink}  
\def\BibTeX{{\rm B\kern-.05em{\sc i\kern-.025em b}\kern-.08em
    T\kern-.1667em\lower.7ex\hbox{E}\kern-.125emX}}

\renewcommand{\_}{\textunderscore\allowbreak} 

\usepackage{caption}
\begin{document}

\title{Hardening the OSv Unikernel with Efficient Address Randomization: Design and Performance Evaluation}

\author{\IEEEauthorblockN{Alex Wollman\,\orcidlink{0009-0000-6260-2750}}
\IEEEauthorblockA{\textit{The Beacom College of Computer and Cyber Sciences} \\
\textit{Dakota State University}\\
Madison, SD, USA \\
alex.wollman@dsu.edu}
\and
\IEEEauthorblockN{John Hastings\,\orcidlink{0000-0003-0871-3622}}
\IEEEauthorblockA{\textit{The Beacom College of Computer and Cyber Sciences} \\
\textit{Dakota State University}\\
Madison, SD, USA \\
john.hastings@dsu.edu}
}
\maketitle

\begin{abstract}

Unikernels are single-purpose library operating systems that run the kernel and application in one address space, but often omit security mitigations such as address space layout randomization (ASLR). In OSv, boot, program loading, and thread creation select largely deterministic addresses, leading to near-identical layouts across instances and more repeatable exploitation. To reduce layout predictability, this research introduces ASLR-style diversity into OSv by randomizing the application base and thread stack regions through targeted changes to core memory-management and loading routines. The implementation adds minimal complexity while preserving OSv's lightweight design goals. Evaluation against an unmodified baseline finds comparable boot time, application runtime, and memory usage. Analysis indicates that the generated addresses exhibit a uniform distribution. These results show that layout-randomization defenses can be efficiently and effectively integrated into OSv unikernels, improving resistance to reliable exploitation.
\end{abstract}


\section{Introduction}
\label{intro}
In the past decade, the utilization of container technologies (Linux Containers (LXC), Docker, Podman) \cite{nider_comparison_nodate, pahl_cloud_2019, watada_emerging_2019} has taken center stage in the cloud environment \cite{zhang_comparative_2018, nizar_analysis_2024, pahl_cloud_2019}. These technologies offer isolated environments and reduced resource requirements \cite{zhang_comparative_2018, pahl_cloud_2019} compared to the full operating system emulation offered by virtual machines (VMs) \cite{zhang_comparative_2018}. These containers require fewer resources than VMs, allowing more containers to operate in the same space a single VM once occupied. As the demand for cloud resources continues to increase, the last decade saw the emergence of a technology called unikernels \cite{madhavapeddy_unikernels_2013-1, el_ioini_unikernels_2023}. Although themselves inspired by library operating systems \cite{madhavapeddy_unikernels_2013, engler_exokernel_1995, porter_rethinking_2011}, unikernels are single-purpose applications that combine, or `uni'fy, the kernel and application into one binary \cite{wollman_survey_2024, madhavapeddy_unikernels_2013-1}. The result is a single address space binary that is compiled only with what it needs to execute.

A binary that is compiled with only the bare essentials may, at first glance, appear largely innocuous. A web server, for example, does not need audio drivers, word processing capabilities, user account management, or even user interface devices. Removing these capabilities and all support therein removes a tremendous amount of unnecessary code. In unikernels, however, such redactions often extend beyond the removal of superfluous functionality and exclude critical security mechanisms such as address space layout randomization (ASLR) and stack canaries \cite{wollman_survey_2024, michaels_ncc_group-assessing_unikernel_securitypdf_2019, wollman_ceker_2024, talbot_security_2020}. 

The absence of these protections is frequently justified by arguments that assert a reduced attack surface \cite{madhavapeddy_unikernels_2013, rapp_analysis_nodate, compastie_unikernel-based_2018, wollman_ceker_2024, wollman_survey_2024} or a lack of terminal-like applications \cite{michaels_ncc_group-assessing_unikernel_securitypdf_2019, talbot_security_2020, wollman_ceker_2024, wollman_survey_2024, bratterud_enhancing_2017}. Regarding the first claim, there is a notable lack of empirical research proving the relationship between reduced functionality and a measurably smaller attack surface \cite{rapp_analysis_nodate, wollman_survey_2024}. Precisely defining ``attack surface" itself remains a non-trivial and under-specified problem in the literature: this problem extends beyond the unikernel literature \cite{theisen_attack_2018}. 

Regarding the latter claim, while the absence of terminal-like applications may constrain certain proof-of-concept attacks, such limitations do not fundamentally preclude exploitation. Return-oriented programming utilizes chunks of code called     `gadgets' to exploit a system and gain control without the need of a terminal \cite{hiser_ilr_2012, smashingTheGadgets}. With the rate of vulnerability discovery rising year after year, a security lapse has the potential to be a devastating problem \cite{heffley_can_2004}.

Traditional operating system security mechanisms such as ASLR are absent from the OSv unikernel by default, creating a fundamentally weaker security posture. The lack of address randomization results in deterministic and largely identical memory layouts across OSv instances. This predictability substantially reduces the technical barrier to exploitation and enables simpler and more reliable attack techniques. With existing unikernel research that focuses primarily on performance over security \cite{wollman_survey_2024}, the goal of this research is to introduce address randomization in the OSv unikernel and investigate the impacts it has on the performance of the unikernel. 
As such, this research addresses the following research questions:
\begin{enumerate}[label={\textbf{RQ\arabic*:}},left=1.0em]
    \item What set of mechanisms can effectively implement address randomization in OSv? 
    \item What is the impact of address randomization on OSv boot time, application runtime, and memory usage? 
    \item Do the randomization mechanisms generate uniformly distributed values? 
    \end{enumerate}

\section{Design Goals \& Approach}
\label{design}
\subsection{Methodology}

This work adopts a Design Science methodology and employs a single-case mechanism experiment \cite{DesignScienceMethodology} in which a single component is altered and the entire system is monitored so that observed effects can be attributed to that change. In this case, the mechanism under test is OSv's address-selection behavior for key memory regions during boot, program loading, and thread creation. 

In this paper, the term `address randomization' is used specifically to refer to ASLR-style randomization of OSv’s memory regions. This research modifies how memory addresses are generated by introducing ASLR-style diversity utilizing a pseudo-random number generator (PRNG) to produce page-aligned randomized addresses within OSv-specific bounds of the form \texttt{0x0000XXXXXX000000}. 

The artifact generated by this research is an OSv unikernel with ASLR-like capabilities. Performance metrics of unikernel boot time, application execution time, and memory usage are collected for comparison with the original OSv. These changes are evaluated to see whether they achieve similar performance metrics (boot time, application execution time, and memory usage) while adding memory layout diversity.

The development approach utilizes as many native language features and built-in capabilities of OSv to minimize the amount of new code introduced to the code base. This approach intends to retain the minimalistic design theory of unikernels, while achieving the necessary randomization capabilities. The primary languages utilized by OSv are C and C++, which natively support a wide range of randomization capabilities. 

\subsection{System and Experimental Setup}
OSv is the unikernel utilized in this research, using commit hash 0xf515d191fcefe9dc2aa8c544849ac4554ce1fecd pushed on Sunday April 27, 2025. The device utilized to perform the benchmarking is a Dell Latitude 7450 with 250GB hard drive, 16GB RAM, and an Intel Core Ultra 7 165U. VMware Workstation 17 Pro build 17.6.2 is the hypervisor used to build the analysis VM. The analysis VM is a 64-bit Ubuntu 24.04.3 LTS using the Linux kernel 6.14.0-36 generic, built with 4GB RAM, 100GB hard drive, and 2 processors with 2 cores each. 

Once built, OSv uses QEMU to run. Although OSv supports a build tool similar to Docker, the scripting suite, stored in the \texttt{scripts} folder, was utilized instead. The benchmarking tool utilized by this research required additional system configuration steps, which are detailed in Appendix \ref{appendix_a}.

\section{Background: OSv Boot Path and Deterministic Address Selection}
\label{background}

OSv's default memory layout is largely determined by deterministic address choices made during boot, program loading, and thread stack allocation. The following overview traces the specific control paths and functions that establish these mappings, providing the context needed to understand the implementation changes presented in this research.

OSv supports both x64 (Intel) and aarch64 (ARM) architectures. OSv utilizes a script that, amongst other things, identifies the platform's architecture and begins the unikernel's build configuration. This configuration is utilized by a Makefile at the base of the code repository containing a target called \texttt{stage1\_targets}. One dependency of \texttt{stage1\_targets}, for this research specifically, is the 64-bit version of \texttt{boot.S}. While a full discussion of \texttt{boot.S} is outside the scope of this research, it is important to note the unikernel build and boot process begins here. The relevant component is its invocation of \texttt{main}, which eventually calls \texttt{main\_cont}, both of which are defined in \texttt{loader.cc}. \texttt{main\_cont} is responsible for running most, if not all, of the unikernel's startup systems: ACPI, SMP, BSD, VFS, IRQ, networking, etc. 

Near the beginning of \texttt{main\_cont} is where \texttt{elf::create\_main\_program} is called. This is a misleadingly simple function that instantiates the \texttt{elf::program} class, which by default uses the hard-coded base address of 0x0000100000000000. The \texttt{elf::program} constructor invokes the \texttt{load\_segments} function, which loads the segments of the target application using the \texttt{load\_segment} function. Within this function, the \texttt{map\_anon} function is called, which is responsible for memory allocations. If no memory address is provided, which is the default behavior, memory is allocated beginning at the hard-coded address of 0x0000200000000000.

Near the end of \texttt{main\_cont} is where the thread is created that is responsible for executing the application itself. First \texttt{pthread\_attr\_init} is called which initializes a \texttt{thread\_attr} structure, responsible for storing the stack pointer and other vital information for the thread. After \texttt{pthread\_attr\_init} returns, the structure is passed to \texttt{pthread\_create} along with the function \texttt{do\_main\_thread}. At the end of \texttt{pthread\_create} a new \texttt{pthread} instance is created that accepts both \texttt{do\_main\_thread} and \texttt{thread\_attr} structure as arguments.\footnote{The structure is assigned values throughout the process, but nothing that affects the design or evaluation of this research.} The \texttt{pthread} constructor uses a constructor list to initialize its member variables, which includes a function call to \texttt{attributes} which uses the \texttt{thread\_attr} structure. Within this function, the \texttt{allocate\_stack} function is called, which itself calls the \texttt{map\_anon} function to allocate the requested amount of memory for the stack: the \texttt{thread\_attr.stack\_size} contains this value. Of note, the \texttt{map\_anon} function's first argument is the requested starting address, which in this function is \texttt{NULL}. Because it is \texttt{NULL}, \texttt{map\_anon}'s call to \texttt{allocate} starts the search for an appropriate memory address at the hard coded value of 0x0000200000000000 as discussed above.

These locations, program loading and stack allocation, motivate the design decisions discussed next in Section \ref{results}.

\section{Implementation Results}
\label{results}

Address randomization was successfully introduced to OSv by modifying the memory management components of the unikernel, resulting in randomization of the stack and base application addresses. These modifications involved 93 lines of code split primarily between two files. Minor modifications were also required to elements such as function declarations and definitions to support the new additions\footnote{In total 8 files were modified: core/app.cc, core/elf.cc, core/mmu.cc, include/osv/mmu-defs.hh, libc/libc.cc, libc/pthread.cc, loader.cc, bsd/cddl/contrib/opensolaris/head/thread.h.}. These implementation results, along with the discussion that follows, address \textbf{RQ1}.

\subsection{Randomization Primitives and Constraints}
Two sets of constraints are used when building the random numbers: \texttt{BIT\_CHECK} and \texttt{ELF\_RND\_MASK} in \texttt{elf.cc} and \texttt{BIT\_CHECK} and \texttt{STACK\_RND\_MASK} in \texttt{mmu.cc}. In the first instance, \texttt{BIT\_CHECK} and \texttt{ELF\_RND\_MASK} are set to 0x0000100000000000 and 0x00001fffff000000 respectively, and in the second 0x0000300000000000 and 0x00003fffff000000. The values are chosen to set the lower and upper bounds of the addresses, respectively, as well as clearing the lower 3-bytes for page alignment and program usage.

Three functions implemented by this research form the foundational primitives for address randomization: \texttt{rand\_gen}, \texttt{check\_rdrand\_support}, and \texttt{seed\_generator}. The random number generator is implemented in the \texttt{seed\_generator} function. This function uses a built-in feature of the C++ library called \texttt{random\_device} that generates a 32-bit number. In order to generate a 64-bit number the function is called twice: the first value is left shifted and then bitwise XOR'd with the second value.

The \texttt{check\_rdrand\_support} function is a wrapper for the \texttt{\_\_get\_cpuid} function. This function queries the CPU directly and returns a bitmask. Right shifting this mask by 30 provides access to the bit indicating if \texttt{rdrand} is supported (bit is set to 1) or not (bit is set to 0.) 

If \texttt{check\_rdrand\_support} returns 0, then the process described in Section \ref{background} is followed using default values, otherwise the \texttt{seed\_generator} function is called to retrieve a value. The value is ANDed with the \texttt{BIT\_CHECK} constraint to ensure a proper minimum value, and is repeated until such a value is returned. Once a value is found, it is ANDed with \texttt{ELF\_RND\_MASK} or \texttt{STACK\_RND\_MASK} as detailed above to set the appropriate upper bound.

\subsection{Program Base Randomization}
\label{elfSection}
In the OSv unikernel, randomizing the program base causes \texttt{main} to be randomized, as well as any global variables declared by the application. Within the \texttt{elf::create\_main\_program} function, if \texttt{rand\_gen} returns an address, it is used; otherwise, a default value is used as described in Section \ref{background}.

\subsection{Heap Behavior}
Unlike traditional binaries, the heap in OSv is created and loaded as a segment. The randomized base address is used as the starting address and each segment is loaded at increasing offsets of 0x1000. In this way, the heap is not independently randomized, but is derived from the (randomized) base address.

\subsection{Stack Randomization and pthread Attributes}
\label{mmu}
A new boolean variable was added to the \texttt{thread\_attr} structure called \texttt{random\_stack}. When this value is true, the new enum value of \texttt{mmap\_rand} (enum defined in \texttt{mmu-defs.hh}) is OR'd with the existing \texttt{stack\_flags} value to trigger address randomization. A second constructor was added to the \texttt{thread\_attr} structure to accept a boolean value that populates the new variable called \texttt{random\_stack}. The function responsible for triggering this sequence of events is \texttt{pthread\_attr\_init}, and is modified to accept the new boolean value and pass it to the newly defined constructor.

The stack address randomization process follows the same sequence of function calls as detailed in Section \ref{background}, but a boolean flag attribute is added to the \texttt{thread\_attr} structure to enable the \texttt{mmu\_rand} flag in \texttt{allocate\_stack}, triggering the address randomization in \texttt{map\_anon}.

Similar to the program base process described above, \texttt{map\_anon} has been modified to implement the same \texttt{rand\_gen} functionality if the \texttt{mmu\_rand} flag is set. Importantly, a random address will be generated and used regardless of the address provided.

\section{Evaluation}
The benchmark tool (detailed in Appendix \ref{appendix_a}) was utilized to collect different metrics including application run time, memory usage\footnote{Memory usage is measured in Mebibytes (MiB) which represents $2^{20}$ bytes.}, and boot time. The two instances of OSv utilized were the base OSv unikernel, with commit hash f515d191, and the modified unikernel created as a result of this research. The benchmark tool was executed 303 times per unikernel for an approximate total of 600 runs\footnote{Seven results were unusable due to errors in the output.}

Table \ref{modifiedAndUnmodifiedSDTable} shows the standard deviation (SD) for three different metrics: total run time (measured in milliseconds), unikernel boot time (measured in milliseconds), and memory usage (measured in mebibytes) for the modified and unmodified unikernel. The benchmark tool calculates the unikernel boot time value by recording the current time, invoking a function to boot the unikernel, and upon return recording the time again. These two values are then subtracted to determine the boot time. The same start time value is used when determining total run time, only a new end time is retrieved once the unikernel execution has completed.

Due to challenges in distinguishing memory usage between the Virtual Memory Manager and the unikernel, the total memory usage metric calculated by the benchmark tool consists of both values. This limitation is detailed by the benchmarking tool's documentation. The full quote is provided in Appendix \ref{appendix_b}.

To determine whether the observed differences were statistically significant, Levene's test was utilized. The results showed that there were no statistically significant differences in Run Time or Boot Time in Runs 2 and 3, with Run 2 Boot Time p-value of 0.0717 and Run Time of 0.4391 and Run 3 Boot Time p-value of 0.4234 and Run Time 0.3239. Run 2 had no statistically significant difference in Memory Usage with a p-value of 0.1902, addressing \textbf{RQ2}.

\begin{table}[htbp]
\caption{Modified (M) and Unmodified (UM) Unikernel SD}
\label{modifiedAndUnmodifiedSDTable}
\begin{center}
\begin{tabular}{|c|c|c|c|}
\hline
Run&\textbf{Run Time (ms)}&\textbf{Boot Time (ms)}&\textbf{Mem Usage (MiB)} \\
\hline
M 1 & $\pm$105.7 & $\pm$55.6 & $\pm$1.6 \\
\hline
UM 1 & $\pm$199.7 & $\pm$166.2 & $\pm$5.5 \\
\hline
M 2 & $\pm$61.4 & $\pm$36.1 & $\pm$1.0 \\
\hline
UM 2 & $\pm$61.2 & $\pm$27.9 & $\pm$0.9 \\
\hline
M 3 & $\pm$147.4 & $\pm$41.2 & $\pm$1.5 \\
\hline
UM 3 & $\pm$113.9 & $\pm$39.3 & $\pm$1.0 \\
\hline
\end{tabular}
\label{tab1}
\end{center}
\end{table}

Tables \ref{unmodifiedCITable} \& \ref{modifiedCITable} show the 95\% confidence intervals for the same three metrics, with the left value being the lower confidence interval and the right the upper.

\begin{table}[htbp]
\caption{Unmodified Unikernel 95\% Confidence Intervals}
\label{unmodifiedCITable}
\begin{center}
\begin{tabular}{|c|c|c|c|c|c|c|}
\hline
Run&\multicolumn{2}{|c}{\textbf{Run Time (ms)}}&\multicolumn{2}{|c|}{\textbf{Boot Time (ms)}}&\multicolumn{2}{|c|}{\textbf{Mem Usage (MiB)}} \\
\hline
1 & 9290.8 & 9370.0 & 773.0 & 819.1 & 126.6 & 128.8 \\
\hline
2 & 8932.5 & 8956.8 & 623.7 & 634.6 & 135.4 & 135.8 \\
\hline
3 & 9052.4 & 9097.9 & 640.7 & 656.4 & 132.9 & 133.3 \\
\hline
\end{tabular}
\end{center}
\end{table}

\begin{table}[htbp]
\caption{Modified Unikernel 95\% Confidence Intervals}
\label{modifiedCITable}
\begin{center}
\begin{tabular}{|c|c|c|c|c|c|c|}
\hline
Run&\multicolumn{2}{|c}{\textbf{Run Time (ms)}}&\multicolumn{2}{|c|}{\textbf{Boot Time (ms)}}&\multicolumn{2}{|c|}{\textbf{Mem Usage (MiB)}} \\
\hline
1 & 9145.5 & 9187.7 & 679.1 & 701.3 & 133.4 & 134.1 \\
\hline
2 & 8993.1 & 9017.5 & 655.7 & 670.0 & 135.6 & 136.0 \\
\hline
3 & 9087.6 & 9147.7 & 657.1 & 673.9 & 134.0 & 134.6 \\
\hline
\end{tabular}
\end{center}
\end{table}

Beyond performance impact, the generated addresses were also evaluated to see if they exhibited a uniform distribution. The distributions of the randomized numbers were calculated based on the region: addresses for the main address, heap, and stack. Using the Kolmogorov–Smirnov test, uniformity was evaluated at ($\alpha=0.05$) and because (n$>$50), the asymptotic critical value ($D_{crit}=1.36/\sqrt{n}$) \cite{kolmogorovSmirnovTest} was used. The results failed to reject the NULL hypothesis, indicating the data likely comes from a uniform distribution. This addresses \textbf{RQ3}.

\section{Limitations \& Future Work}
\label{futureWork}

Several alternative designs were evaluated but rejected due to complications. One approach generated a pseudo-random number in the \texttt{thread} constructor and passed it to \texttt{init\_stack}. However, altering the stack pointer at that stage did not change the allocation, only the address, which is certain to lead to a variety of memory corruption bugs. Other attempts experimented with different PRNGs, but were ultimately not successful due to unsafe register usage (in-line assembly) or call timing (functions not available when needed).

The benchmarking tool revealed that allocating large chunks of memory (10MB) invokes the `malloc\_large' function; a separate branching path that does not result in the generation of randomized addresses. As the name implies, this function is responsible for handling very large memory allocation requests. Due to the requested memory size and the potential impact randomizing that amount of space has on the system, it is instead left for future work. Although this was the only path discovered during the course of this research, more research is needed to discover if other paths exist and what triggers them. 

Another avenue for future work is the randomization of library addresses. These non-randomized addresses pose a threat to the unikernel if a malicious actor were to discover an initial attack vector. Access to static memory makes exploits more reliable, and since the libraries are not randomized, developing a sophisticated exploit would be easier to achieve. Randomizing the starting address of a library means the addresses of its functions are not consistent across devices\footnote{The functions will still have a consistent offset relative to the starting address of the library.}. Information leaks and additional exploitation steps are needed to retrieve these base addresses, complicating the exploit development process. 

Another avenue for future work is re-evaluating the heap implementation. Due to how OSv is built, the heap address is loaded as a segment and thus is related to the main program's address. In other kernels, a separate execution path is responsible for allocating and assigning the heap, completely separate from segments. Future work in this area could be to remove the segment and treat the heap as a separate component, managed independently or by the memory management unit. Then address randomization could be handled using the technique presented in this research.

Another avenue for future work is to expand or use different PRNG techniques. This research focuses solely on x86 and Intel processors for the support of \texttt{rdrand}. Early attempts at platform independent techniques utilized custom PRNGs, but were discarded in favor of built-in randomization. The challenge with custom PRNGs is to generate a seed value that does not require randomization itself. Using an improper seed value will result in deterministic or easily guessable random numbers. Future work could re-investigate custom PRNGs and seed generation to remove the \texttt{rdrand} requirement. Other research could implement similar processor-specific capabilities for different architectures and processors.

Finally, another avenue for future work is applying the contributions of this paper to other unikernels. The techniques used were specifically developed to utilize built-in features of the C/C++ languages, making them easily transferable. The \texttt{check\_rdrand\_support} function may need modification depending on the underlying architecture, but this change should be minimal.

\section{Related Work}
\label{relatedWork}

Previous work has examined unikernel security from survey and system perspectives. Several studies catalog the security properties of unikernels and compare them to more conventional deployment models, highlighting both reduced attack surfaces and new classes of risk \cite{duncan_enhancing_2016, verstraete_how_2021, wollman_ceker_2024}. Independent assessments and industry reports similarly point out that unikernels are not inherently secure and require additional hardening to withstand modern attack techniques \cite{michaels_ncc_group-assessing_unikernel_securitypdf_2019, talbot_security_2020}. These works establish the need for explicit mitigations, but do not provide or evaluate concrete implementations of ASLR within production unikernel systems.

Other researchers have proposed mechanisms to strengthen unikernels using techniques that differ fundamentally from this research. Hardware-assisted designs use Intel SGX to isolate unikernel components and protect sensitive data from a compromised operating system or hypervisor \cite{shen_occlum_2020, sfyrakis_uniguard_2018, oleksenko_varys_2018}. Other work leverages Intel Memory Protection Keys (MPK) to enforce intra-unikernel isolation, partitioning a single image into protection domains with distinct access rights \cite{sung_intra-unikernel_2020}. These approaches tackle the same broad problem of hardening unikernel deployments, but rely on specialized hardware features and more intrusive changes to the execution model. In contrast, this research explores how classic address randomization can be integrated into an existing unikernel (OSv) with minimal modifications and without assuming trusted hardware beyond commodity CPU support for randomness.

More recently, researchers have adapted ASLR-like techniques to constrained or specialized environments. Function-level randomization has been proposed for microcontroller-class devices, using TrustZone-M and memory protection hardware to introduce diversity into embedded firmware layouts \cite{luo_faslr_2022}. Other efforts focus on hardening Linux container deployments through namespace configuration, capability reduction, and integration of traditional kernel-level mitigations \cite{jian_defense_2017}. These systems address different platforms (IoT devices and containerized workloads) but share the goal of bringing established exploit mitigations into new deployment models. This research achieves this goal by incorporating address randomization into a mature unikernel implementation with minimal impact on performance and size.

ASLR and related address randomization techniques have a long history in general-purpose operating systems. They are widely used to prevent control-flow hijacking and data-oriented attacks by making the location of code and data segments unpredictable \cite{smashingTheGadgets}. At the same time, studies have demonstrated ways to weaken or bypass ASLR, for example, by exploiting side channels, partial information leakage, or micro-architectural behavior \cite{smashingTheGadgets, evtyushkin_jump_2016}. This body of work informs the design choices in this research, treating address randomization as a best-effort mitigation whose effectiveness depends on the size and quality of the randomized region, the entropy available, and the practical constraints imposed by the platform.

Taken together, previous work demonstrates that unikernels require explicit hardening and that address randomization remains a relevant defense across diverse environments. This research's contribution bridges these threads: providing a concrete case study of integrating ASLR-style defenses into OSv, quantifying their cost in a realistic setting, and analyzing the benefits and limitations of such an approach for unikernel-based deployments.
\section{Conclusion}
\label{conclusion}
Unikernels offer significant advantages for cloud and edge deployments, yet their lack of built-in exploit mitigations presents a critical security concern. This work demonstrates that address randomization can be effectively integrated into the OSv unikernel with minimal code changes; preserving its lightweight design principles. Empirical results using Levene's test show no statistically significant impact on boot time, execution time, or memory usage, confirming that security enhancements need not compromise performance. Additionally, statistical analysis using the Kolmogorov-Smirnov test verifies that the resulting memory layouts exhibit a uniform distribution, improving resistance to exploitation. These findings indicate that traditional operating system mitigations can be practically adopted within unikernel environments, providing a viable path toward more secure unikernel-based systems.
\printbibliography

\appendices
\section{}
\label{appendix_a}
The benchmarking tool used is found on the GitHub repository \nolinkurl{https://github.com/Genez-io/unikernel-benchmark}. The benchmarking tool uses Docker to build the unikernels, which introduces a challenge of nesting virtual devices. To nest virtualization technologies on Windows, the Device Guard Readiness tool\footnote{https://www.microsoft.com/en-us/download/details.aspx?id=53337} was used to \textit{remove} Device Guard and Hyper-V. This is necessary for VMware Workstation as the two hypervisor platforms conflict when attempting to nest virtualization technologies. Once disabled by using the flag '-disable' the device must be rebooted and changes accepted at the prompts. Then the two virtualization settings can be selected in the VM Settings (Virtualize IOMMU and Virtualize Intel VT-X/EPT or AMD-V/RVI) and the VM booted to run the benchmarking tool. This step is not necessary if no benchmarking is desired. Minor modifications were then necessary to point the tool at the appropriate GitHub repository to pull and build the modified unikernel.

\section{}
\label{appendix_b}
``To measure memory consumption, we opted to measure the total memory occupied by both the VMM and the unikernel. This decision was made due to the lack of a straightforward and universal method to determine which memory pages are utilized by the VMM and which are utilized by the unikernel. Consequently, it was not possible to accurately determine the standalone memory consumption of the unikernel. To monitor the memory usage of a process, we developed a Python class that tracks the RSS memory of a given process ID and records the measurements at regular intervals. The memory readings are stored in a list for further analysis and reference.''\footnote{\nolinkurl{https://github.com/Genez-io/unikernel-benchmark/blob/main/benchmark-tool/README.md\#memory-measurement}}

\end{document}